\documentclass[manuscript=article]{achemso}

\author{Frank M. Abel}
\affiliation{The Volgenau Department of Physics, United States Naval Academy, Annapolis, Maryland 21402, United States}
\author{Jaehyung Lee}
\affiliation{Department of Materials Science and Engineering, Whiting School of Engineering, Johns Hopkins University, Baltimore, Maryland 21218, United States}
\author{Charles R. Campbell}
\affiliation{Department of Materials Science and Engineering, Whiting School of Engineering, Johns Hopkins University, Baltimore, Maryland 21218, United States}
\author{Kamal Choudhary}
\email{kchoudh2@jhu.edu}
\affiliation{Department of Materials Science and Engineering, Whiting School of Engineering, Johns Hopkins University, Baltimore, Maryland 21218, United States}
\affiliation{Department of Electrical and Computer Engineering, Whiting School of Engineering, Johns Hopkins University, Baltimore, Maryland 21218, United States}

\title[RamanGPT]{RamanGPT: Bidirectional Mapping Between Crystal Structures and Raman Spectra with Graph Neural Networks and Generative Transformers}

\usepackage{amsmath,amssymb}
\usepackage{graphicx}
\usepackage{booktabs}
\usepackage{xcolor}
\usepackage{url}

\begin{document}

% =====================================================================
%  Abstract 
% =====================================================================
\begin{abstract}
Raman spectroscopy is one of the most accessible vibrational probes in materials laboratories, but its forward problem (structure to spectrum) is bottlenecked by the cost of density functional perturbation theory, and its inverse problem (spectrum to structure) typically relies on retrieval against curated references. We introduce RamanGPT, a deep-learning framework that addresses both directions for crystalline inorganic materials. The forward model, an Atomistic Line Graph Neural Network (ALIGNN), is trained on the 5{,}099-material Computational Raman Database and predicts 200-bin spectra over 50-1000~cm$^{-1}$ with 42.5\% having a cosine similarity greater than or equal to 0.354 suggesting qualitative features of the target spectrum. The model also shows some qualitative agreement with the approximate features and appearance of similar relative intensity of the modes to an experimental measurement of metallic 1T VSe$_{2}$, a system absent from the training set. The inverse model fine-tunes a large language model via Quantized Low-Rank Adaptation on Raman-plus-formula prompts, recovering lattice parameters with mean absolute errors of 1.14-2.16~\AA{} and reduced-formula consistency of 86.8\% on 508 held-out materials. A cosine-similarity matcher and an inverse$\rightarrow$relax$\rightarrow$forward consistency loop are deployed at \url{https://atomgpt.org/raman}.
\end{abstract}

% =====================================================================
%  Main text 
% =====================================================================
\section{Introduction}

Raman scattering, discovered in 1928,\cite{raman1928} has become the most widely used vibrational probe in materials science because it gives information on chemical bonding, crystal symmetry, strain, and phase content non-destructively, with approximately micron spatial resolution and modest instrumentation cost.\cite{long2002,schrader2008,ferrari2013,zhang2016} The technique is now routine across batteries and solid electrolytes,\cite{famprikis2019} electrocatalysis,\cite{zhan2021} biomedical imaging,\cite{abramczyk2013} and planetary exploration,\cite{beegle2015,lopezreyes2013, veberanda2023} and underpins flagship instruments on Mars 2020 and the ExoMars rover. Yet two complementary computational tasks remain difficult. The forward task-predicting a Raman spectrum from a known crystal structure requires phonon eigenmodes at $\Gamma$ together with derivatives of the electronic susceptibility along each Raman-active normal-mode coordinate. In density functional perturbation theory (DFPT), this scales as $3N+1$ self-consistent calculations per material with $N$ atoms in the unit cell,\cite{baroni2001,porezag1996,bagheri2023} restricting high-throughput screens to a few thousand compounds.\cite{liang2019,bagheri2023,li2024raman,taghizadeh2020} The inverse task, inferring a structure from a measured spectrum, is harder still: spectral features couple to the structure through the dynamical matrix and Raman tensor in a non-linear, multi-step fashion,\cite{reichardt2019} chemically distinct materials can yield similar peaks, and broadening, anharmonicity, and instrumental factors further degrade the mapping.

The traditional remedy for the inverse problem is retrieval against curated references. Experimental libraries include RRUFF,\cite{lafuente2015} the KnowItAll Raman collection,\cite{knowitall} and the Raman Open Database.\cite{elmendili2019} Computational counterparts include the Computational Raman Database (CRD; 5{,}099 DFPT spectra),\cite{bagheri2023} the hybrid-functional set of Li \textit{et al.},\cite{li2024raman} the ab-initio 2D-materials libraries of Taghizadeh \textit{et al.}\cite{taghizadeh2020} and Li \textit{et al.},\cite{li2025ht2d} the WURM mineral collection,\cite{caracas2011} and the early high-throughput study of Liang \textit{et al.}\cite{liang2019} Matching software such as the ExoMars Raman pipelines have operationalized this strategy.\cite{hermosilla2012} Retrieval is fast and interpretable, but cannot be generalized.

Machine learning (ML) offers a path beyond retrieval. In the forward direction, graph neural networks (GNNs) treat a crystal as a graph of atoms and bonds, and have set state-of-the-art accuracy on dozens of solid-state property prediction benchmarks,\cite{xie2018,schutt2018,batzner20223,choudhary2021alignn,choudhary2022dl,schmidt2019} with extensions to the phonon density of states,\cite{gurunathan2023,kong2022,chen2021} electronic DOS,\cite{kaundinya2022} and X-ray absorption spectra.\cite{rankine2020,carbone2020} The Atomistic Line Graph Neural
Network (ALIGNN)\cite{choudhary2021alignn} is particularly well suited to vibrational targets because it explicitly encodes both bond distances and bond-angle triplets through a paired crystal graph and line graph; both quantities directly determine the dynamical matrix and the polarizability derivatives that control Raman activity.

In the inverse direction, ML has historically been cast as classification-mapping a spectrum to one of a fixed set of mineral labels.\cite{liu2017,sang2022,berlanga2022,ibtehaz2023} While such convolutional neural networks (CNNs) exceed 95\% top-1 accuracy on RRUFF, the answer is constrained to the training labels and provides no atomic coordinates. To predict an
actual structure, an inherently generative model is required. Recent work has therefore re-cast crystal structure prediction as a sequence-generation problem: if a structure is serialized into text (lattice constants, angles, element symbols, fractional coordinates), a transformer trained on millions of such serializations can learn the joint distribution of plausible crystals.\cite{flam2023language,gruver2024fine,antunes2024crystal,rubungo2023llm,zeni2023mattergen} The AtomGPT framework\cite{choudhary2024atomgpt} adapts this idea to materials design by fine-tuning instruction-tuned large language models (LLMs) such as Mistral-7B-Instruct\cite{jiang2023mistral} on Alpaca-style prompts that pair a serialized crystal with a target property. Two recent extensions are directly relevant here: DiffractGPT,\cite{choudhary2025diffractgpt} which recovers lattice parameters and atomic positions from powder X-ray diffraction (PXRD), and MicroscopyGPT,\cite{choudhary2025microscopygpt} which infers 2D-material structures from STEM images. Both rely on parameter-efficient Quantized Low-Rank Adaptation (QLoRA),\cite{hu2021lora,dettmers2023qlora} which trains $\sim$0.3\% of the parameters while leaving the pretrained weights frozen.

Raman spectra, however, present a less direct inverse problem than PXRD or STEM. PXRD peak positions are directly related to lattice spacings through Bragg's law and STEM directly images the lattice; in both cases, the measurement is tied to structural quantities by short causal chains. Raman spectra, by contrast, encode vibrational information through
the full force-constant network and the response of the electronic susceptibility to atomic displacements,\cite{bagheri2023,porezag1996,baroni2001}. This adds layers of indirection between measurement and the material's structure, making Raman highly sensitive to bonding, defects, strain, and local effects, but generally less uniquely invertible without reference spectra or first-principle modeling. Whether an LLM can learn to invert this relationship with useful accuracy is an open question.

\begin{figure*}[t]
\centering
\includegraphics[width=0.95\linewidth]{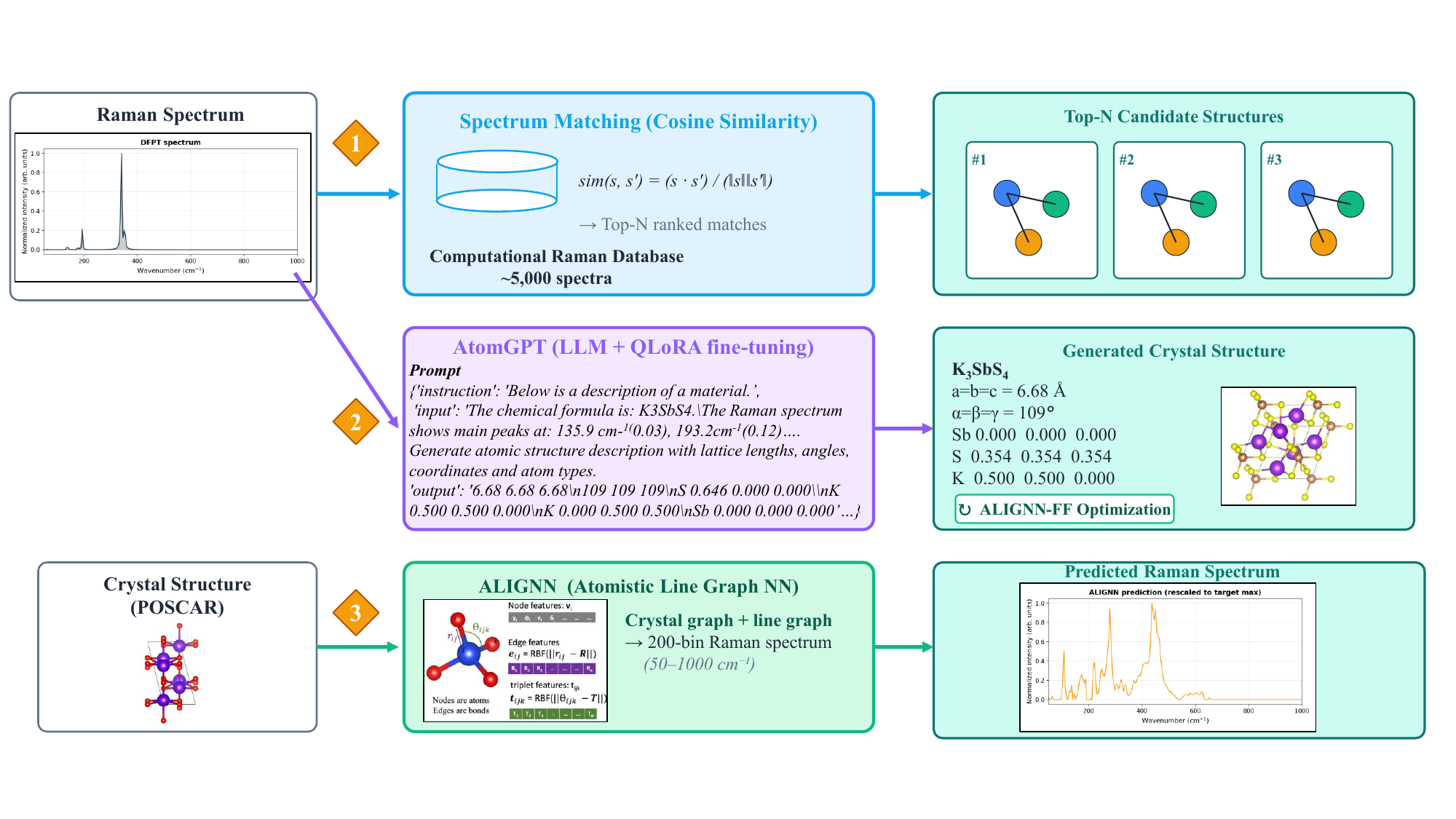}
\caption{Overview of the RamanGPT framework (1) Cosine-similarity spectrum matching against the Computational Raman Database (CRD; 5{,}099 DFPT spectra)\protect\cite{bagheri2023} returns top-$N$ candidate structures. (2) AtomGPT (Mistral-7B-Instruct fine-tuned with QLoRA) generates a complete crystal description-lattice lengths, angles, elements, and fractional coordinates from the spectrum and an explicit chemical formula. The input for (1) and (2) is shown as a DFPT spectrum, but could be an experimental measurement. (3) ALIGNN consumes a POSCAR-format crystal and predicts a 200-bin Raman spectrum over 50-1000~cm$^{-1}$. The web application additionally supports an inverse$\rightarrow$ALIGNN-FF$\rightarrow$forward consistency workflow, in which a generated structure is relaxed and re-fed to the forward model. All three modules are deployed at \protect\url{https://atomgpt.org/raman}.}
\label{fig:schematic}
\end{figure*}

Here we introduce \textbf{RamanGPT}, a unified framework that addresses the forward, inverse, and matching tasks for crystalline Raman spectroscopy in a single deployed system (Figure~\ref{fig:schematic}). The forward module is an ALIGNN multi-output regressor; the inverse module is a QLoRA-fine-tuned Mistral-7B-Instruct generator; the third module performs cosine-similarity retrieval against the CRD with configurable Gaussian broadening and chemical-formula filtering. All three are accessible from a Raman Suite web application at \url{https://atomgpt.org/raman}, which also exposes an inverse$\rightarrow$relax$\rightarrow$forward consistency workflow in which a candidate structure produced by the inverse model is relaxed with the ALIGNN-FF universal force field\cite{choudhary2023alignnff} and re-fed to the forward model for self-consistency checking.

\section{Results and Discussion}

We use the CRD,\cite{bagheri2023} which contains DFPT-computed Raman tensors for 5{,}099 materials. The CRD builds on the Phonon Database,\cite{togo2015} which provides VASP PBEsol\cite{kresse1996} optimized structures, force-constant matrices, and phonon eigenvectors, and is linked to the Materials Project\cite{jain2013} via material identifiers. Materials in the CRD are pre-screened for Raman activity, dynamical and thermodynamic stability ($E_{\text{hull}}<0.1$~eV/atom), and a band gap exceeding 0.5~eV; the database spans oxides (52\%), halides (27\%), sulfides, phosphates, silicates and carbonates, with 2-100$+$ atoms per primitive cell across all seven crystal systems. For polycrystalline samples, the orientation-averaged Raman intensity is $I_{\text{Raman}} = 45\bar{a}^{2} + 7\bar{\gamma}^{2}$, where $\bar{a}$ and $\bar{\gamma}^{2}$ are the isotropic and anisotropic invariants of the Raman tensor.\cite{bagheri2023,long2002,porezag1996} Spectra were discretized into 200 bins (4.75~cm$^{-1}$ resolution) over 50-1000~cm$^{-1}$ and normalized to peak intensity. We used an
80/10/10 train/validation/test split, yielding 509 forward and 508 inverse test materials.

The forward model uses the standard ALIGNN architecture\cite{choudhary2021alignn} with four ALIGNN layers and four edge-gated convolution layers (256 hidden features), 92 atomic input features, a 8.0~\AA{} cutoff with 12 neighbors, and a 200-feature regression head. We trained for 100 epochs with AdamW (learning rate 0.001, one-cycle scheduler) and MSE loss at batch size 16. The inverse model fine-tunes Mistral-7B-Instruct\cite{jiang2023mistral} through QLoRA\cite{hu2021lora,dettmers2023qlora} in 4-bit precision, modifying only $\sim$0.3\% of the parameters. The Alpaca-style prompt template is: \textit{``Below is a description of a material. The chemical formula is [formula]. The Raman spectrum is [intensities]. Generate atomic structure description with lattice lengths, angles, coordinates and atom types.''} The output text is parsed into lattice constants (\AA{}), angles (deg), element symbols, and fractional
coordinates. Reduced-formula and space-group analysis is performed through the \texttt{jarvis.core.atoms} canonicalizer and \texttt{spglib}\cite{togo2024spglib} (default symmetry tolerance). Because the chemical formula is supplied in the prompt-matching, the explicit-formula condition of DiffractGPT\cite{choudhary2025diffractgpt} (the formula-match metric reported below) quantifies how reliably the LLM preserves and canonicalizes the input formula through structure generation rather than how often it infers it from the spectrum alone, a distinction we revisit in the comparison below. The matcher applies a configurable Gaussian kernel (default $\sigma\!=\!8$~cm$^{-1}$, matching the broadening used in the CRD reference plots\cite{bagheri2023}) to the input frequency-intensity pairs, resamples onto the CRD grid, and ranks all entries by cosine similarity with optional formula-based pre-filtering.

\begin{figure*}[t]
\centering
\includegraphics[width=0.95\linewidth]{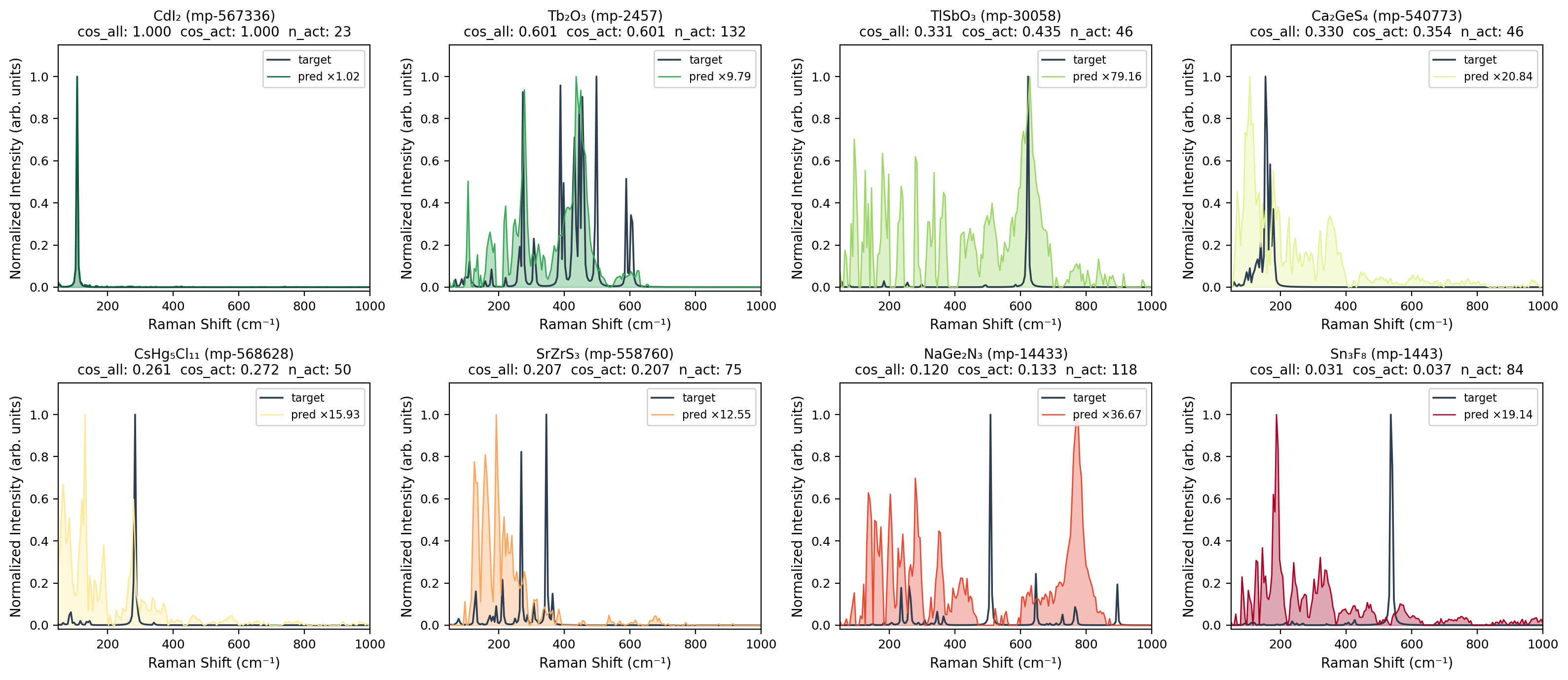}
\caption{Representative forward Raman spectrum predictions generated using the ALIGNN Raman model across varying similarity regimes.
Each panel compares the DFPT-computed target Raman spectrum (dark curve) with the corresponding ALIGNN-predicted spectrum (colored curve) for selected materials from the test set. The eight examples are sampled approximately uniformly from the ranked distribution of prediction quality, spanning the full range from highest to lowest cosine similarity. The metric \texttt{cos\_all} denotes the cosine similarity computed across all 200 Raman bins, while \texttt{cos\_act} is computed only over spectrally active bins where the target intensity exceeds 0.001. The quantity \texttt{n\_act} indicates the number of active target bins. To improve visualization of low-intensity predictions, the predicted spectra were rescaled by the multiplicative factor shown in each legend.}
\label{fig:forward_gallery}
\end{figure*}

\textit{Forward model performance.} Across 509 test materials, the ALIGNN forward model attains a mean MAE of 0.032 with a median MAE of 0.029. Approximately 88\% of predicted spectra fall below MAE $= 0.05$ while 99.4\% remain below MAE $= 0.10$, suggesting a strong overall agreement between predicted and DFPT-computed spectra. However, we note that this MAE is relatively high, with a near-perfect match having an MAE of 0.001, 32x smaller than the mean. Due to this, we consider the cosine similarity as a better metric for determining the quality of spectrum matching. Figure~\ref{fig:forward_gallery} presents eight representative examples sampled uniformly across the ranked distribution of cosine similarity scores, spanning the range from highest to lowest quality predictions. Examining these eight spectra, we observe that a cosine similarity of 0.354 and 0.435 shows peak predictions matching some of the main peaks in the target. However, there seems to be a large number of predicted peaks not present in the target. In comparison, a cosine similarity of 0.601 generally matches the overall features of the target spectrum. Considering the overall cosine similarity distribution, 42.5\% have a cosine similarity of 0.354 or greater, which appears to be around the limit of reasonable matching. Meanwhile, 14.2\% score 0.601 or higher, and only 0.2\% show a near-perfect match. For materials with relatively simple Raman responses, such as CdI$_2$, the model reproduces both peak positions and relative intensities with near quantitative accuracy. As spectral complexity increases, the network continues to recover the dominant vibrational features and overall spectral envelope, although closely spaced sharp peaks are progressively broadened or averaged. Frequency bins reveal that the largest deviations occur within the 100--400~cm$^{-1}$ region, corresponding to the frequency range where most materials in the CRD dataset exhibit strong Raman activity. Prediction errors decrease substantially above 600~cm$^{-1}$ as the spectra become increasingly sparse. This behavior is consistent with previous observations in ALIGNN-based phonon density-of-states prediction studies.\cite{gurunathan2023} The primary failure modes arise from (i) materials containing numerous sharp and densely packed vibrational peaks distributed across the full spectral range, which the model tends to smooth, and (ii) light-element compounds with significant Raman activity extending beyond the 1000~cm$^{-1}$ training window, where truncation limits the model’s ability to fully capture the spectrum.

We further studied how spectral discretization affects accuracy by sweeping the bin count from 50 to 1000 (Table~\ref{tab:bins}). The MAE decreases monotonically from 0.053 (50 bins) to 0.031 (500 bins, a 42\% reduction) and then plateaus, while the success rate (\%~$<$~0.05) jumps sharply from 51\% at 50 bins to 88\% at 200 bins and is flat beyond. We adopt 200 bins as the operating point: it captures 97\% of the best achievable MAE while requiring 2.5-5$\times$ fewer output features than 500/1000-bin variants. Pearson correlation and cosine similarity remain stable across resolutions (0.45-0.47 and 0.53-0.56, respectively), suggesting they are dominated by the inherent sparsity of Raman spectra rather than by discretization.

\begin{table}[t]
\small
\centering
\caption{Effect of spectral bin resolution on the ALIGNN forward model (509 test materials).}
\label{tab:bins}
\begin{tabular}{rcccc}
\toprule
Bins & MAE & \%$<$0.05 & \%$<$0.10 & Peak MAE (cm$^{-1}$) \\
\midrule
50   & 0.053 & 51.1 & 95.7 & 115 \\
100  & 0.038 & 80.0 & 99.0 & 118 \\
\textbf{200} & \textbf{0.032} & \textbf{88.0} & \textbf{99.4} & \textbf{102} \\
500  & 0.031 & 88.6 & 99.4 & 104 \\
1000 & 0.031 & 88.4 & 99.6 & 101 \\
\bottomrule
\end{tabular}
\end{table}

\begin{figure*}[t]
\centering
\includegraphics[width=0.95\linewidth]{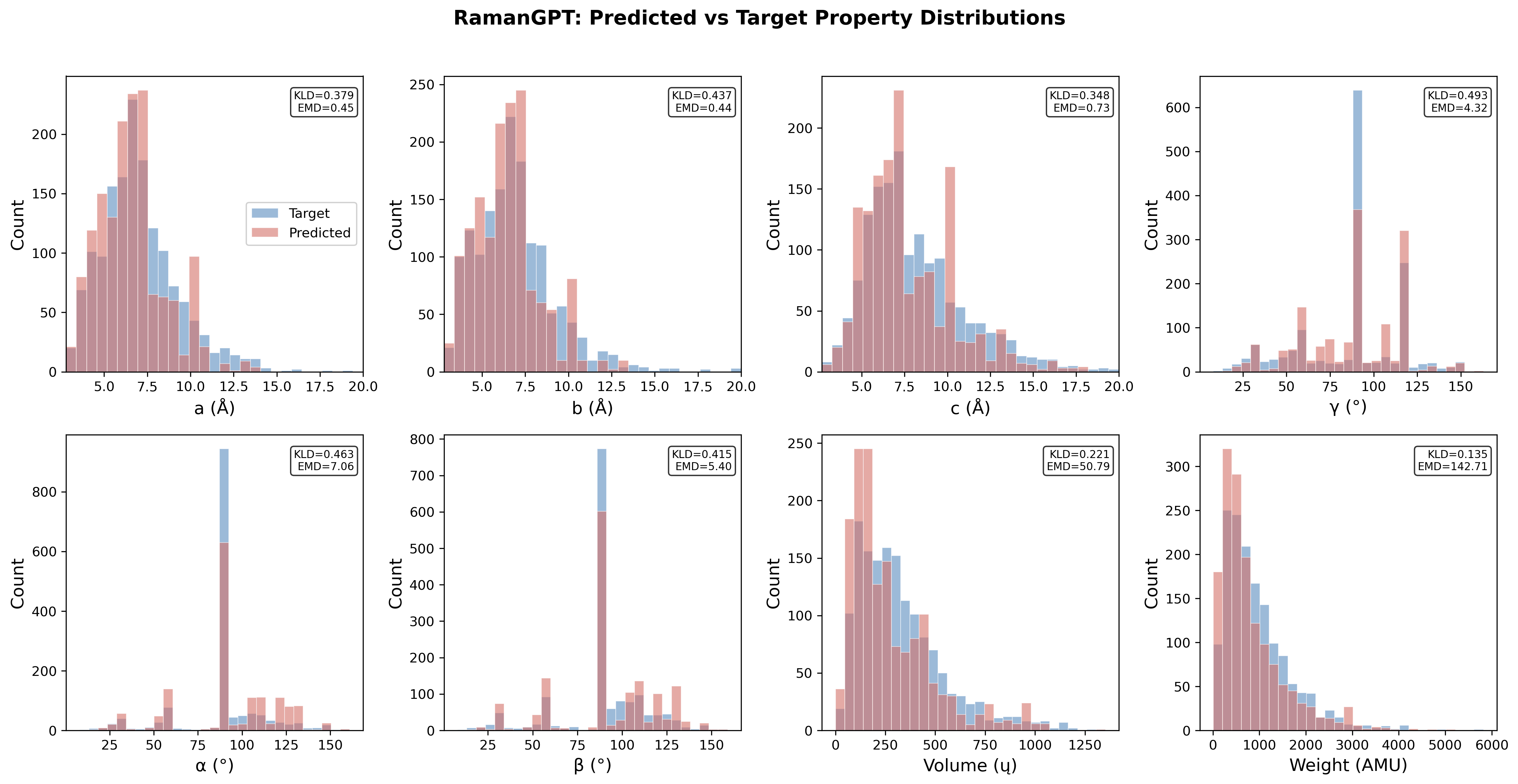}
\caption{Predicted (red) versus target (blue) distributions for the inverse RamanGPT model on 508 held-out materials. Panels show lattice parameters $a$, $b$, $c$ (\AA{}); lattice angles $\alpha$, $\beta$, $\gamma$ g(deg); cell volume (\AA$^{3}$) and atomic weight (AMU). KLD and EMD values are annotated.}
\label{fig:inverse_dist}
\end{figure*}

\textit{Inverse model performance.} The RamanGPT inverse model was evaluated on the 508-material held-out test split. Figure~\ref{fig:inverse_dist} compares predicted and target distributions for the eight key structural quantities, with Kullback-Leibler divergence (KLD) and Earth mover's distance (EMD) annotated; lower values indicate better agreement.\cite{choudhary2025microscopygpt} Table~\ref{tab:inverse} summarises the corresponding mean absolute errors. Lattice-parameter MAEs are 1.14~\AA{} ($a$), 1.20~\AA{} ($b$) and 2.16~\AA{} ($c$). The density distribution shows the closest agreement with the target (KLD~$=$~0.025, EMD~$=$~0.21), suggesting the model captures the overall ``heaviness'' of the unit cell encoded in low-frequency modes. The reduced chemical formula is preserved in 86.8\% of cases, with mean element-composition accuracy of 76.3\%. Even when the exact reduced formula is not preserved, the predicted compositions remain chemically similar.

\begin{table}[t]
\small
\centering
\caption{Inverse model metrics on 508 test materials. KLD and EMD quantify distributional agreement.}
\label{tab:inverse}
\begin{tabular}{lccccc}
\toprule
Property & Min & Max & MAE & KLD & EMD \\
\midrule
$a$ (\AA{})            & 2.90  & 27.49 & 1.137  & 0.379 & 0.447 \\
$b$ (\AA{})            & 2.90  & 27.49 & 1.197  & 0.437 & 0.438 \\
$c$ (\AA{})            & 2.69  & 75.39 & 2.158  & 0.348 & 0.728 \\
$\alpha$ (deg)         & 8.0   & 159.0 & 17.085 & 0.463 & 7.064 \\
$\beta$ (deg)          & 8.0   & 158.0 & 17.590 & 0.415 & 5.40  \\
$\gamma$ (deg)         & 8.0   & 159.0 & 20.803 & 0.493 & 4.317 \\
Volume (\AA$^{3}$)     & 20.66 & 1321  & 107.06 & 0.221 & 50.794 \\
Density (g/cm$^{3}$)   & 0.71  & 11.89 & 0.741  & 0.025 & 0.206 \\
\bottomrule
\end{tabular}
\end{table}

Lattice-angle predictions are noticeably weaker than lattice lengths (MAE 17-21$^{\circ}$). Two factors drive this: most CRD entries are cubic, tetragonal or orthorhombic and therefore have angles clamped at $90^{\circ}$, biasing the LLM toward orthogonal predictions. Raman spectra are more directly sensitive to bond lengths and force constants than to the angular geometry of the unit cell. The space-group match rate (computed via \texttt{spglib} with default symmetry tolerance) is 18.9\%, low in absolute terms but reflective of the extreme sensitivity of Wyckoff assignment to small coordinate perturbations. The median fractional-coordinate RMSE is 0.265, which corresponds to roughly 1.5-3~\AA{} of position error for typical CRD primitive cells (5-12~\AA{} edge length). This is sufficient for identifying the structural motif, but not for crystallographic refinement without post-processing.

The natural baseline for the generative inverse model is the spectrum-matching module of Figure~\ref{fig:schematic}, evaluated on the same 508 test spectra: nearest-neighbor cosine matching against the remainder of the CRD recovers the correct reduced formula in 41\% of cases and the correct space group in 9\%, with no facility for extrapolation outside the database. The fine-tuned LLM therefore roughly doubles formula-consistency and space-group recovery relative to retrieval, while additionally producing full atomic coordinates ready for downstream relaxation. Compared to DiffractGPT, which is the closest direct analogue and shares the same Mistral+QLoRA backbone, the RamanGPT lattice-parameter MAEs (1.14-2.16~\AA{}) are larger by a factor of 2-7 (DiffractGPT reports 0.17-0.32~\AA{} on the JARVIS-DFT test set under the explicit-formula condition). This gap is the expected price of inverting a vibrational rather than a diffraction signal. PXRD peak positions encode lattice spacings directly through Bragg's law, whereas Raman peaks encode them only indirectly through the dynamical matrix and Raman-tensor response. The 86.8\% reduced-formula match rate is interpretable in the same light. Because the formula is supplied in the prompt, this number quantifies how reliably the LLM preserves and canonicalizes the input formula through structure generation rather than infers it from the Raman trace alone, with the residual 13.2\% explained predominantly by hallucinated atom counts when generating large or chemically unfamiliar unit cells. A formula-free ablation (and element-list conditioning, analogous to DiffractGPT's three-scenario study) is left for future work.

\begin{figure}[t]
\centering
\includegraphics[width=\linewidth]{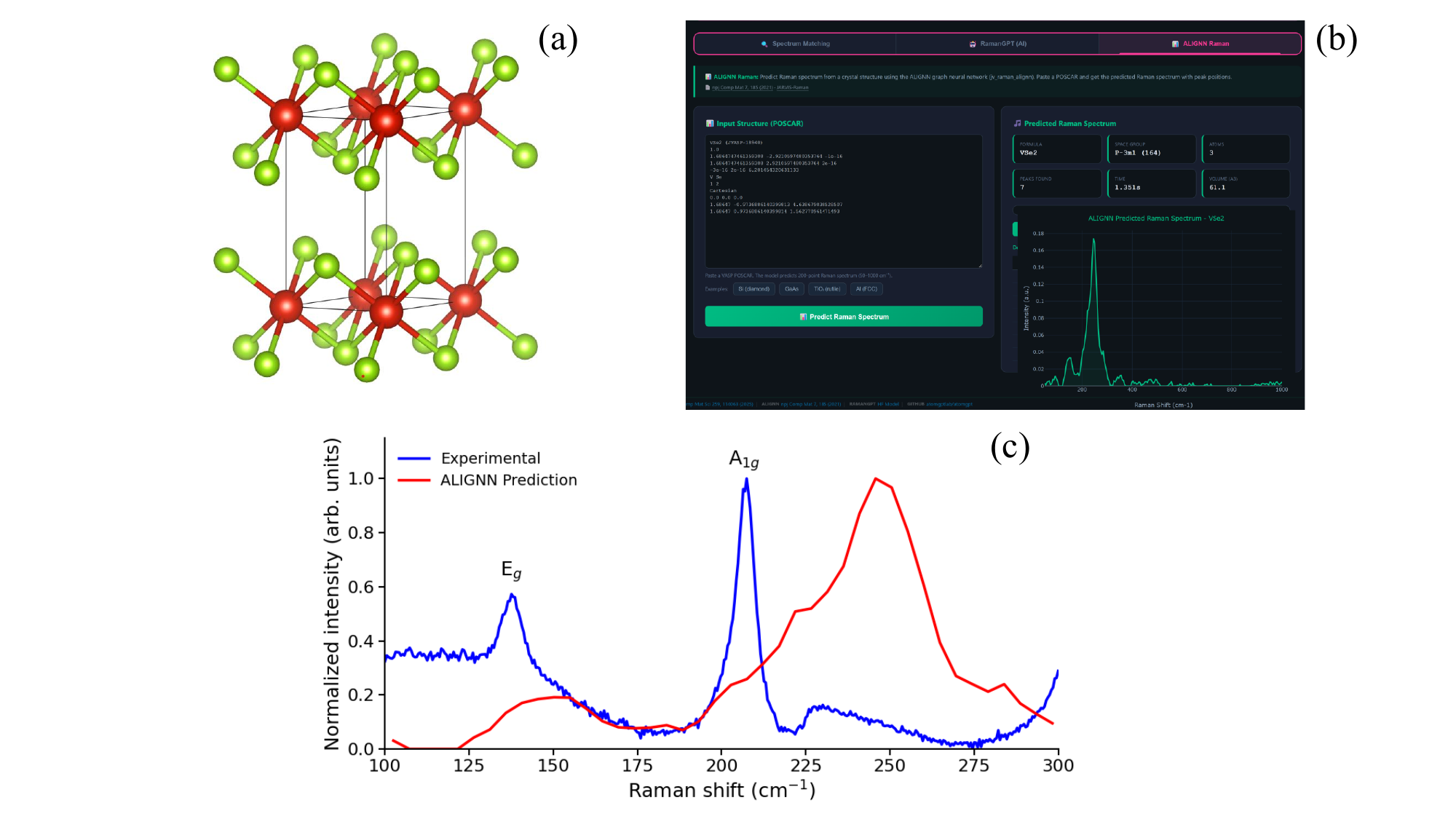}
\caption{Forward-model comparison to experimentally measured VSe$_{2}$ in the ALIGNN-Raman web app. (a) Known JVASP-18940 VSe$_{2}$ structure used as POSCAR input. (b) ALIGNN forward prediction from the input structure, displayed in the web app. (c) Normalized ALGINN prediction overlay with normalized experimental Raman spectrum showing qualitative agreement. The experimental spectrum dataset was obtained from previously published work \cite{wines2025quantum}.}
\label{fig:vse2}
\end{figure}

\textit{Forward-model comparison to experiment measurement.}
Figure~\ref{fig:vse2} shows the ALIGNN forward module applied to VSe$_{2}$ within the ALIGNN-Raman web app, as a deployment-stage out-of-distribution test against laboratory data. Raman spectrum data of an exfoliated approximately 1.7 nm 1T VSe$_{2}$ flake from previously published work was used for model comparison, which was measured to have two modes at 206.6 $\pm$ 0.4 cm$^{-1}$ and 137.4 $\pm$ 0.3cm$^{-1}$, identified as an A$_{1g}$ and E$_g$ respectively. Further details can be found here \cite{wines2025quantum}. The known JVASP-18940 VSe$_{2}$ structure (panel a) was supplied as a POSCAR to the ALIGNN forward model, which returned the predicted spectrum (panel b) directly within the web interface. The experimental and ALGINN predictions are plotted together (panel c), showing qualitative agreement in terms of similar features. The prediction shows the main features (modes) shifted to higher $cm^{-1}$, but the main features appear to have similar relative intensities to the experimental data. Additionally, VSe$_{2}$ is absent from the training set: in the 1T phase, VSe$_{2}$ is metallic\cite{bonilla2018vse2} and therefore does not satisfy the band-gap pre-screen of the CRD \cite{bagheri2023}, so no VSe$_{2}$ entry was seen during training. The forward model nonetheless reproduces the dominant experimental features for a metallic layered chalcogenide, which is encouraging for guiding experimentalists in identifying where to look for modes of materials when DFT results or database comparisons are not available. We treat this as a qualitative demonstration rather than a quantitative metallic-systems benchmark, which awaits a dedicated training set. The companion inverse$\rightarrow$ALIGNN-FF$\rightarrow$forward consistency workflow described above is presented separately within the Raman Suite interface.

\begin{figure}[t]
\centering
\includegraphics[width=\linewidth]{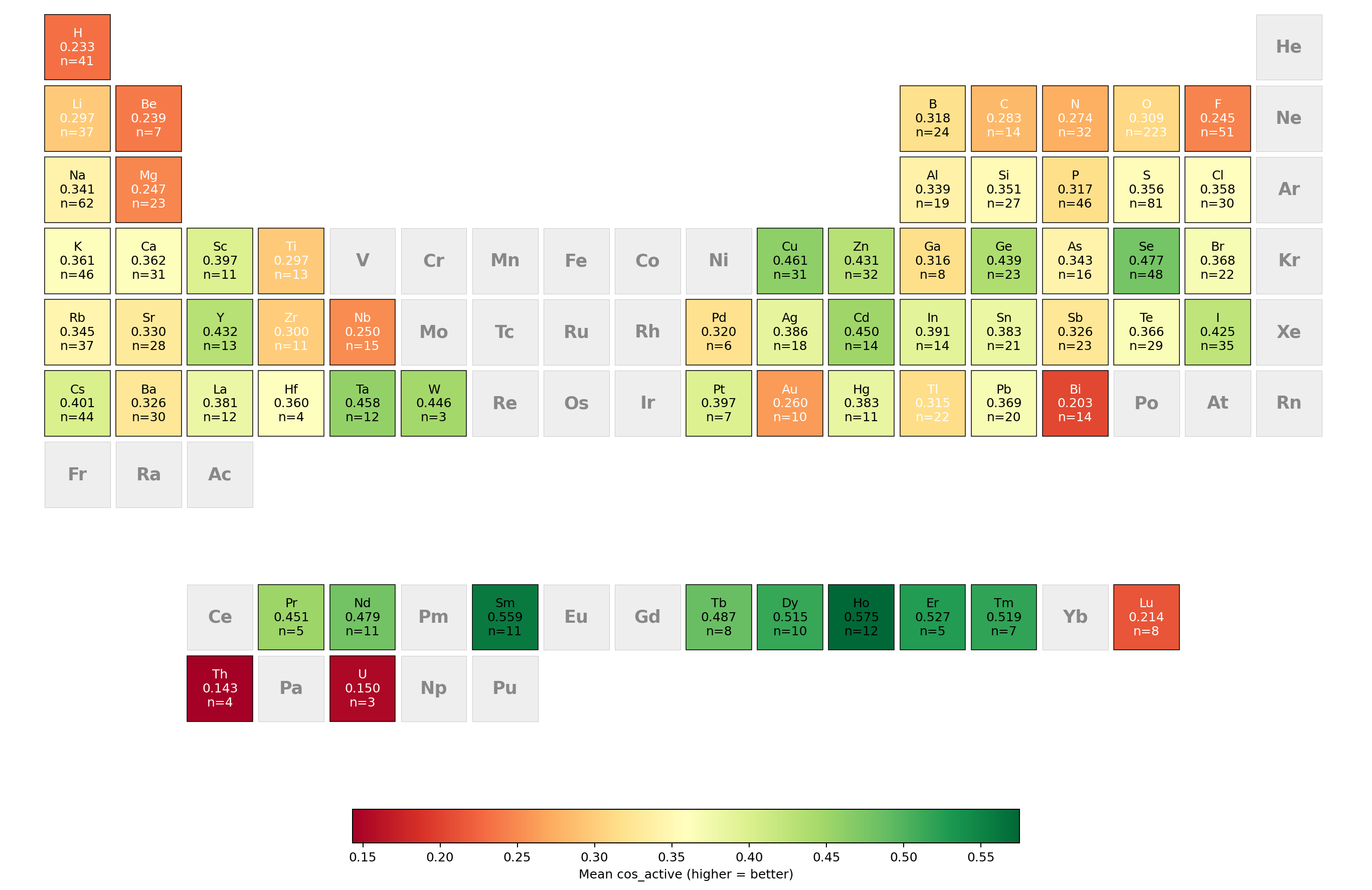}
\caption{Element-wise performance trends of the ALIGNN Raman forward model across the test dataset. The periodic table displays the mean active-region cosine similarity (\texttt{cos\_active}) computed over all test compounds containing each element. Each tile reports the average cosine similarity together with the number of associated test compounds ($n$). Elements shown in gray correspond to cases with fewer than three compounds in the test set and are excluded from trend interpretation.}
\label{fig:trend}
\end{figure}

\textit{Element-wise performance trends.} Figure~\ref{fig:trend} shows the mean active-region cosine similarity grouped by elemental composition across the test dataset, providing insight into how model performance varies throughout chemical space. The highest-performing regions are concentrated among several lanthanides, chalcogens, and late transition-metal systems, where Raman responses are often dominated by relatively localized and well-defined vibrational modes. In particular, rare-earth-containing compounds exhibit consistently high cosine similarity values, suggesting that the model effectively captures their characteristic low-frequency vibrational behavior. Moderate performance is observed across many alkali, alkaline-earth, and post-transition-metal compounds, indicating broad transferability of the learned representations. Interestingly, lower similarity values are more common for light-element materials and actinide-containing compounds, which frequently exhibit broader spectral distributions, dense multi-peak structures, or vibrational activity extending beyond the 50--1000~cm$^{-1}$ prediction window used during training. Variability in performance also partially reflects dataset imbalance, as several elements are represented by only a small number of compounds in the test set. Overall, the observed periodic trends demonstrate that the ALIGNN Raman model generalizes across diverse chemistries while revealing systematic dependencies between spectral complexity and prediction accuracy.

\textit{Outlook.} RamanGPT establishes that crystalline Raman spectroscopy admits an end-to-end deep-learning treatment in both directions. The forward question, whether a graph network can, in some cases, reproduce DFPT-quality Raman spectra at high throughput, is answered by showing a cosine similarity greater than or equal to 0.354 for 42.5 \% of the holdout dataset, together with qualitative comparison to a laboratory-measured metallic chalcogenide outside the training distribution. The inverse question we posed at the outset: whether an LLM can invert the multi-step mapping from vibrational features to atomic positions at useful accuracy, is answered tentatively in the affirmative. With explicit chemical-formula conditioning, a QLoRA-fine-tuned 7B-parameter model recovers lattice constants to within 1.14-2.16~\AA{} and reduces the input formula correctly in 86.8\% of cases, despite the substantially more indirect spectrum-to-structure mapping than for PXRD or STEM. Several limitations point to immediate extensions. First, the training set is restricted to inorganic crystals with band gaps above 0.5~eV and dynamical stability at the $\Gamma$-point;\cite{bagheri2023} extending the model to metals, defective phases and disordered solids will require either resonance-aware physical descriptors\cite{reichardt2019,hashemi2019} or substantially augmented training data. Second, the spectral window of 50-1000~cm$^{-1}$ excludes the high-frequency stretching modes of light-element compounds. Widening the window is a straightforward retraining exercise, but it increases the inverse-model context length. Third, lattice-angle prediction is biased by the strong $90^{\circ}$ prior in the training data, motivating future conditioning on auxiliary inputs such as crystal system, space-group symbol, or Bravais lattice, analogous to the chemical-formula conditioning that improves DiffractGPT.\cite{choudhary2025diffractgpt} A natural follow-up is therefore an ablation of the prompt across no-formula, element-list, and explicit-formula conditions to disentangle spectrum-driven structure recovery from formula-driven canonicalization. Fourth, broader experimental validation against RRUFF\cite{lafuente2015} and additional laboratory measurements remains an active line of work; the integrated round-trip inverse$\rightarrow$ALIGNN-FF$\rightarrow$forward consistency criterion is already exposed through the deployed web application and provides an interpretable confidence signal for inverse predictions. Finally, the same framework should generalize to infrared, terahertz, and inelastic neutron spectra simply by altering the underlying database, and to combine Raman+PXRD or Raman+microscopy inputs by extending the prompt template. By unifying retrieval, forward prediction, and generative inversion in a single open framework, RamanGPT brings the language-model-as-crystallographer paradigm to the most ubiquitous vibrational probe in materials laboratories.

\section{Methods}\label{sec:methods}

RamanGPT is a unified framework that combines generative transformers and graph neural networks to model both forward (structure $\rightarrow$ spectrum) and inverse (spectrum $\rightarrow$ structure) Raman problems. The inverse module builds on the AtomGPT paradigm, a decoder-only transformer architecture~\cite{choudhary2024atomgpt} fine-tuned from
Mistral-7B-Instruct using Quantized Low-Rank Adaptation (QLoRA). Input prompts consist of a chemical formula token concatenated with a discretized Raman spectrum, and the model auto-regressively generates a structured crystal description including lattice parameters, angles, and fractional atomic coordinates. The core transformer operation is scaled dot-product attention,
\begin{equation}
\mathrm{Attention}(Q,K,V) =
\mathrm{softmax}\!\left(\frac{QK^{\!\top}}{\sqrt{d_k}}\right)V,
\end{equation}
where $Q$, $K$, and $V$ are learned projections of token embeddings.

The forward model employs the Atomistic Line Graph Neural Network (ALIGNN)~\cite{choudhary2021alignn}, which represents a crystal using both an atomic graph $\mathcal{G}$ (atoms as nodes, bonds as edges) and its corresponding line graph $\mathcal{L}(\mathcal{G})$ to encode bond-angle interactions. Node and edge features are updated through interleaved message passing,
\begin{equation}
\mathbf{h}_i^{(l+1)} = \phi\!\left(\mathbf{h}_i^{(l)},
\sum_{j \in \mathcal{N}(i)} \psi\!\left(\mathbf{h}_i^{(l)},
\mathbf{e}_{ij}^{(l)}, \mathbf{h}_j^{(l)}\right)\right),
\end{equation}
enabling accurate prediction of vibrational spectra. The model outputs a fixed-length Raman spectrum discretized over the target frequency range.

To improve the physical plausibility of generated structures, optional post-processing is performed using ALIGNN-FF~\cite{choudhary2023alignnff}, a universal machine learning force field trained on thousands of DFT energy and force evaluations. This relaxation step reduces geometric artifacts and ensures stability prior to forward spectral validation.

All components are integrated into a single pipeline that supports (i) direct spectrum prediction from structure (ALIGNN), (ii) generative structure prediction from Raman spectra (AtomGPT), and (iii) an optional inverse$\rightarrow$relax$\rightarrow$forward consistency loop using ALIGNN-FF and ALIGNN for validation.

\textit{Code and data availability.} Code, trained models and analysis
scripts: \url{https://github.com/atomgptlab/atomgpt} and \url{https://github.com/atomgptlab/alignn}. Web application
and database: \url{https://atomgpt.org/raman} and \url{https://huggingface.co/datasets/knc6/ramangpt_dft}.

\section*{Author Information}

\textbf{Corresponding Author.} Kamal Choudhary -- Department of Materials Science and Engineering and Department of Electrical and Computer Engineering, Whiting School of Engineering, Johns Hopkins University, Baltimore, Maryland 21218, United States; ORCID: 0000-0001-9737-8074; Email: kchoudh2@jhu.edu.
\\
\textbf{Authors.} Frank M. Abel -- The Volgenau Department of Physics, United States Naval Academy, Annapolis, Maryland 21402, United States; ORCID: 0000-0002-8828-7864.

Jaehyung Lee -- Department of Materials Science and Engineering, Whiting School of Engineering, Johns Hopkins University, Baltimore, Maryland 21218, United States; ORCID: 0009-0005-4280-9144.

Charles R. Campbell -- Department of Materials Science and Engineering, Whiting School of Engineering, Johns Hopkins University, Baltimore, Maryland 21218, United States; ORCID: 0009-0006-1520-0939.

\textit{Notes.} The authors declare no competing financial interest. The views expressed in this article are those of the authors and do not reflect the official policy or position of the U.S. Naval Academy, Department of the Navy, the Department of War or the U.S. Government.

% \section*{Acknowledgments}
% The authors thank the JARVIS and AtomGPT user communities for feedback
% on the deployed Raman Suite. K.C.\ acknowledges support from the
% Whiting School of Engineering at Johns Hopkins University.

% =====================================================================
%  Bibliography
% =====================================================================
\bibliography{references}

\end{document}